\documentclass[preprint]{aastex}
\usepackage{emulateapj5}
\usepackage{epsfig}
\begin{document}

\title{
Using a Hipparcos derived HR diagram to limit 
the metallicity scatter of stars in the Hyades
-- Are Stars Polluted?}

\author{A.~C.\ Quillen$^{1,2}$}
\altaffiltext{1}{
Department of Physics and Astronomy, University of Rochester, Rochester, NY 14627; 
aquillen@pas.rochester.edu}
\altaffiltext{2}{Visitor, Physics Department, Technion, Israel Institute 
of Technology, Haifa 32000, Israel}

\begin{abstract}
Hipparcos parallaxes and proper motions have made it possible
to construct Hertzsprung-Russell (HR) diagrams of nearby
clusters with unprecedented accuracy.  The standard deviation 
of high fidelity non-binary non-variable stars about 
a model stellar evolution 
isochrone in the Hyades cluster is about 0.04 magnitudes.
We use this deviation to estimate an upper limit
on the scatter in metallicities in stars in this cluster.
From the gradient of the isochrones evolution in 
the HR diagram we estimate an upper limit for the scatter of metallicities
$\Delta$[Fe/H]$\lesssim 0.03$ dex, a smaller
limit than has been measured previously spectroscopically.
This suggests that stars in open clusters are formed from gas
that is nearly homogeneous in its metallicity.

We consider the hypothesis that processes associated with
planet formation can pollute the  convection zone of stars.
If the position on the HR diagram is insensitive to the metallicity
of the convection zone and atmosphere, then stars which
have very polluted convection zones can be identified
from a comparison between their metallicity
and position on the HR diagram.
Alternatively if the pollution of the star by metals results
in a large change in the position of the star on
the HR diagram in a direction perpendicular to the isochrone, 
then the low scatter of stars in
the Hyades can be used to place constraints on 
quantity of high-Z material that could have polluted the stars.

\end{abstract}

\section{Introduction}

Both spectroscopic studies (e.g. \citealt{sneden,armosky})
and photometric studies (e.g., \citealt{suntzeff,m13,heald,dacosta,buonanno}) 
have shown that the homogeneity of heavy
element abundances in globular clusters is quite small, typically
less than 0.10 dex in [Fe/H]. 
The scatter of stars from model isochrones in 
HR diagrams derived from Hubble Space Telescope (HST) images in cases (such as NGC 6397) 
has been similar in size to the measurement errors which implies that the 
metallicity variation among the stars in a given 
globular cluster must be less than
a few hundredths of a dex (e.g., \citealt{piotto}).
Although high quality HR diagrams of coeval, low metallicity, old globular clusters 
have been constructed, only quite recently has it been possible to construct
HR diagrams of comparable photometric
quality (errors less than a few hundreds of a V band magnitude) 
for younger and nearly solar metallicity coeval systems such 
as open clusters \citep{debruijne}.

Recent spectroscopic studies have established an as yet
unexplained possible connection between the metallicity of a 
star and the existence of short period 
extra solar planets around the parent 
star \citep{gonzalez99, santos00, gonzalezV, gonzalezVI,santos01}.
Compared to stars in the solar
neighborhood, parent stars of short period extra-solar
planets tend to have enhanced metallicities, 
$\Delta[$Fe/H$] \approx 0.2 \pm 0.2$, 
compared to the sun.
However, there is a significant scatter in the observed metallicities;
they range from about [Fe/H]$ \approx -0.4$ to 0.4 \citep{gonzalezVI,santos01}.

There are two possible explanations that account for the enhanced metallicities 
of parent stars with extra solar planets.
One possibility is that
gaseous planets may be rarer around fairly low metallicity stars.
Because the HST study of the globular cluster 47 Tuc 
([Fe/H] $= -0.7$) did not detect eclipses caused by 
planets, short period Jovian planets could be an order of magnitude 
rarer than in the solar neighborhood \citep{gilliland}.  
However the lack of observed eclipses
may instead be a result of the high stellar density in the cluster which
would have caused the disruption of planetary systems \citep{gilliland}.

Alternatively, gaseous planets may be produced
with equal frequency near stars spanning a range of metallicities, 
but the subsequent evolution of the planetary systems increases 
the metallicities of the parent stars.
There is certainly evidence based on abundance analyses 
in our solar system that impacts from bodies in the asteroid
belt or outer solar system 
have changed the surface abundances of the earth and other planets 
(e.g., \citealt{morbidelli,gautier}).
One consequence of the orbital migrating
of giant planets caused by scattering of a disk of planetesimals
\citep{murray}, is that the parent stars are polluted by   
star-impacting planetesimals \citep{quillenholman}.
We infer that
impacts with the sun probably occurred more frequently
and for more massive bodies during the early era
of our solar system.
The transfer of angular momentum via driving of density waves 
into a proto-planetary gaseous disk will 
cause the orbital migration of a planet, and can also cause
a planet to impact the star (e.g., \citealt{lin,trilling}).
One advantage of the orbital migration scenario involving
ejection of planetesimals compared to
that involving density waves driven into a gaseous disk
is that it pollutes the stars with metal rich planetary material 
when they are older and so
have smaller convection zones  (e.g., \citealt{laughlin97})
allowing a smaller amount of metals to cause a larger 
surface metallicity enrichment.  

Because the fraction of mass in the convection zone
depends on the stellar mass, higher mass stars (e.g., F stars) with
smaller convection zones could be more likely than lower
mass stars with larger convection zones (e.g., K stars) to have enhanced
metallicities resulting from pollution caused by planetary material.
However \citet{pinsono} find no such trend among the parent stars of extra
solar planets.  \citet{santos01} find that the distribution
of metallicities for parent stars of extrasolar planets is
inconsistent with a model where these stars have been polluted
by the addition of high-Z material.

Except for a narrow region at $6500 \pm 300K$, known as the Lithium
dip, F stars keep Lithium on their surface
nearly their entire lifetime (see \citealt{balachandran,soderblom}). 
The absence of Li depletion
implies that the convection zone does not mix with the
interior of the star except within the Lithium dip.
For F stars outside the Lithium dip
little mixing takes place and a metallicity enhancement
would last nearly the entire lifetime of the star, however
F stars within the dip would not be expected to show
metallicity enhancements.
\citet{murray01,chaboyer} suggest
that F stars near the Lithium dip, 
tend to have lower metallicities than F stars outside the Lithium dip.

The stellar sample studied by \citet{murray01,chaboyer} and surveyed
for planets draw on the population of stars
in the solar neighborhood and so   
span a large range in ages and metallicities.
To try and decouple the uncertainty caused by the metallicity scatter
resulting from the range of stellar ages in the solar neighborhood 
from that caused by planet formation
we can examine the metallicity scatter in young clusters.
We can assume that the stars in 
a given stellar cluster are the same age and were formed
from gas that is fairly uniform in metallicity.
Stars in the solar neighborhood are expected to have been 
been born in a variety of environments with about 10\% born
in OB associations which become bound open clusters \citep{roberts}.
The low eccentricity orbits of the solar system planets require that 
the birth aggregate of the sun 
was less than a few thousand stars \citep{adams}, and similar
in size to the Trapezium cluster.
Open clusters such as the Hyades, with about 400 known members 
could have been as large as the Trapezium when younger.
Based on the study of \citet{adams}, we expect that short period
extra solar planets should have a high probability of 
surviving disruption from the passage of nearby stars in open clusters.
Despite the fact that most stars in the solar neighborhood
were formed in smaller, unbound groups, since
planetary systems are likely to survive in open clusters, 
we can use observations of the stars in open clusters
to explore the possibility that significant stellar 
metallicity enhancements (pollution) are likely to occur after a star is formed.
If metallicity enhancements
occur, they would occur over a fairly short timescale,
$\sim 10^6$ years for the orbital migration scenario
involving density waves in a gaseous disk \citep{trilling},
and $\sim 10^7$ years
for the migration scenario involving scattering of planetesimals \citep{murray}.
In open clusters, planet systems should survive long enough
that they could have caused metallicity enhancements. 
Furthermore, nearby open clusters have nearly solar metallicities
and so are a better match to the properties of 
parent stars of extra solar planets than are the stars in globular
clusters which are comparatively extremely  metal poor.

It is difficult to envision any planet formation mechanism 
that would {\bf not} cause differing amounts of metallicity 
pollution in different solar systems.  
We therefore expect a scatter in the metallicities
in individual stars in any given stellar cluster.
We might also expect a reduction in [C/Fe] because the inner
solar system is deficient in light elements such as carbon.
However despite earlier reports,
low values of [C/Fe] are not observed in the parent stars
of the short period extra solar planets \citep{gonzalezVI}.
If ice rich cometary material from 
the outer solar system is incorporated into the star
on a later timescale, it is possible that light element abundance could be 
restored.  However this could only occur if material 
from the outer solar system can survive evaporation.
Evaporation rates are higher near the 
high mass and more luminous stars, however these stars, because of their
lower mean densities, are also less likely to cause objects to fragment.
Smaller sized bodies because of their larger surface area are less
likely to achieve final impact, particularly after a series
of close approaches which can be caused
by resonant trapping \citep{quillenholman}.

Spectroscopic studies of F stars in individual open clusters
and moving groups have found that the metallicity scatter
is small $\Delta[$Fe/H] $\lesssim 0.1$ 
\citep{friel92,friel90a,friel90b}.
These authors also measured no measurable variation in the [C/Fe] ratio among
all stars in all clusters studied.
These studies would appear to rule out significant metallicity enhancement
resulting from planet formation and subsequent planetary evolution 
in most stars.  However the number of F stars observed 
in each cluster in these studies was not more than a dozen.  
Establishing cluster
membership is not always unambiguous and so the samples of stars 
chosen for detailed spectroscopic study were not complete.

\section{A Limit in the metallicity scatter from the Hyades cluster HR diagram} 

Hipparcos parallaxes and proper motions have made it possible
to construct Hertzprung-Russell diagrams of nearby
clusters with unprecedented accuracy \citep{debruijne}.
With improved distances and a sample of 92 high fidelity cluster
stars, \citet{debruijne} have constructed an HR diagram with
estimated errors in the log of the luminosity in solar units 
that average 0.030.
The high fidelity sample excludes member beyond 40pc from the cluster
center, multiple stars, variable stars and stars with peculiar or uncertain 
parallaxes (see section 9.3 of \citep{debruijne}).
Stars were {\bf not} excluded based on their position in the HR diagram.

We show in Figure 1 as data points, the difference between the positions of
the stars on the HR diagram and the 630 Myr evolutionary
track at Z=0.024 interpolated from those given
as tables by \cite{girardi}.  The mean difference between
the track and stars is not zero. This implies that the evolutionary
track is not a perfect match to the HR diagram of the cluster.  
It is impressive that the HR diagram is so
good that the evolutionary tracks (and associated bolometric corrections
and color to effective temperature calibrations) can be tested to the 
level of 0.05 in the log of the luminosity.
Nevertheless trends or smooth deviations from zero
on this plot are not important for our study since we are primarily
interested in the scatter of the points.
The ages differences in the stars should be small
compared to the age of the cluster, so other than
observational errors, the 
variations in the stellar metallicities is the primary
factor which would cause a scatter
in the position of the points on the HR diagram about the isochrone.


\includegraphics[angle=0,width=3in]{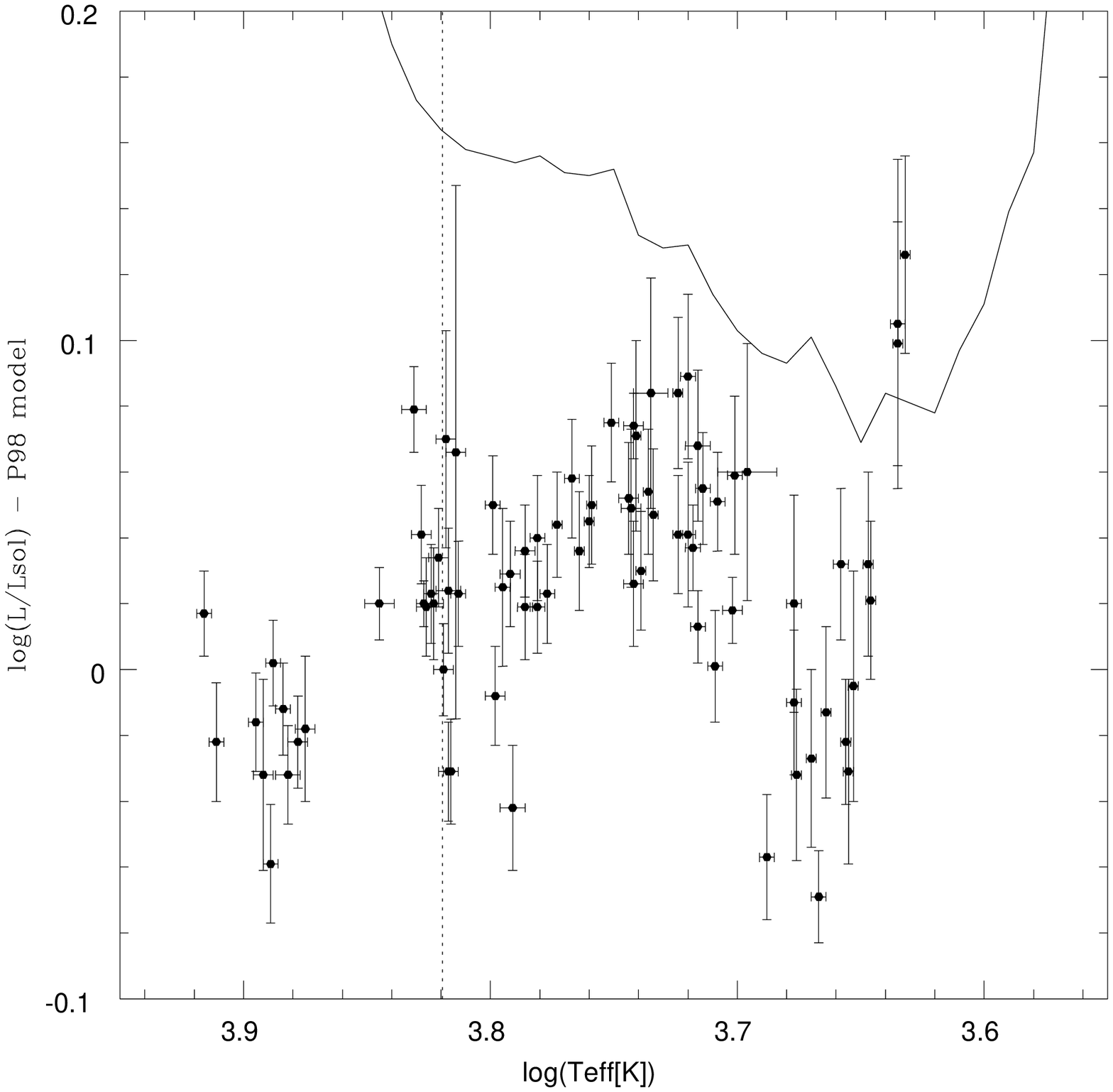}
\begin{quote}
\baselineskip3pt
{\footnotesize Fig.~1--   The points show the difference
between the 92 high fidelity clusters Hyades stars identified
by \citet{debruijne} and the Z=0.024, 630 Myr evolutionary track of 
\citet{girardi}.
We use temperatures and luminosities derived by \cite{debruijne}.
The scatter from the mean has a standard deviation of $\sigma =0.04$
in the log of the luminosity.
We show as a solid line the difference between coeval evolution tracks
that differ by 0.1 dex in [Fe/H].
The dotted line shows the position of the Lithium dip in the Hyades.
}
\end{quote}
\smallskip

We see from Figure 1 that
the scatter off the evolution track is less than 
$\sim 0.05$ in the log of the luminosity compared to the mean local value.
However an evolutionary track that differs by 0.1 dex in [Fe/H] (shown
as the solid line in Figure 1) is offset by 
about 0.15 in the log of the luminosity.  
We therefore estimate that the star to star differences in metallicity
could be only as large as 0.03 dex in [Fe/H].
This only a bit larger that that estimated for the solar
neighborhood by \cite{murray01}, 
0.02 dex corresponding to $0.5 M_\oplus$ of accreted iron from
the inner solar system.
This limit for the standard deviation of 0.03 dex in [Fe/H], 
is better than that previously estimated by \citet{friel90a} from 
less than a dozen F stars which was roughly 0.1 dex.
This low level of scatter in the metallicities suggests that 
the gas from which the stars form is quite homogenous in metallicity.
Because massive stars become supernovae well
before low mass stars are finished accreting this limit also
suggests that nearby supernovae are not capable of 
substantially enriching the ISM in their own nearby star forming regions.

Stellar isochrones are computed assuming that stars do not
accrete additional high-Z material after formation.
Compared to a star which has not accreted additional material, 
where would we expect such a star
to lie on the HR diagram in the cluster?
To know the answer to this question we would need
to calculate additional stellar evolutionary tracks from stellar models.  
While the properties
of the radiative zone might be similar because it
would have the same metallicity, and the energy
transport through the convection should not be strongly
affected by the opacity, the stellar atmosphere would 
be redder and since this sets the boundary condition
for the stellar model, we do not expect the star
to lie at exactly the same position on the HR diagram.

We consider two possibilities:  1) The position in
the HR diagram for a star which has accreted high-Z material
is similar to one that has not.  2) There is a significant
difference in the location on the HR diagram and in a direction
that would distance the star off the isochrone.
If the first possibility is true then
we suggest that spectroscopic analyses can be used
to identify candidate stars which have higher metallicities
than would be expected from their position on the HR diagram.
If such stars are found it would provide strong evidence
that they had been polluted by high-Z material after formation,
and by processes associated with planet formation.

If the second possibility is likely then we emphasize
that the scatter from the isochrone observed
in the HR diagram can be used to constrain stellar pollution scenarios.
Since the level of scatter has a standard deviation
about 0.03 dex in [Fe/H] the average star could
only have accreted about 0.7 $M_\oplus$ of accreted iron from
the inner solar system.

Only a few stars remain outside the local mean in Figure 1 with deviations
greater than 0.05 in the log of the luminosity.   
However stars with short period giant planets are fairly common.
In the solar neighborhood 
6-8\% of solar type stars have short period giant extra solar planets 
\citep{mayor,marcy}.  Of the 92 high fidelity stars we expect 
$6 \pm 2$ with significantly higher metallicities.   
None of the stars is sufficiently off the evolutionary track that
an enhancement of 0.1-0.2 dex could be allowed.  This would
suggest that no star in the Hyades is likely to have metallicities
enhanced by the $30 M_\oplus$ of rocky material needed to
account for the enhanced metallicities of the parent stars of
short period extra solar planets \citep{murray01,laughlin00,chaboyer},
unless these stars happen to lie exactly on the same isochrone
as their non polluted counterparts.

We have checked that there is no correlation between the magnitude
of the scatter in the HR diagram and the amount of lithium
for the 19 F stars in the high fidelity sample
of \cite{debruijne} that have Lithium abundances listed by 
\citet{boesgaard87,thorburn}).  The observed scatter, even
though it seems to peak near the Lithium dip,
does not seem to be obviously related to the mixing below
the convection zone.     We find no relation between
the scatter off the isochrone and the effective temperature of the star
as would be expected from a pollution model which is sensitive to 
the percentage of mass within the convection zone.


\section{Discussion}

In this letter we have used the high quality HR diagram
of the Hyades to place limits on the scatter in
the metallicities of this cluster.  We find that the scatter
in [Fe/H] has standard deviation approximately
0.03 dex.   This estimate is an improvement upon
previous spectroscopic estimates and suggests that open
clusters are formed from gas that is homogenous in metallicity.

We have introduced the possibility
that the extremely low scatter from a predicted isochrone
can be used to constrain stellar pollution scenarios which
involve the addition of high-Z material to the star
from processes associated with planet formation.
This can be done if stellar evolution tracks are calculated
which include the addition of high-Z metals after formation
to see where stars lie on the HR diagram.
If polluted stars lie off the isochrone for unpolluted stars
then the scatter of stars in the HR diagram can be
used to constrain the extent of stellar pollution.
Otherwise comparisons between spectroscopic measured metallicities
and those estimated from the mean properties of the cluster 
can be used to identify
stars which are likely to have been polluted by high-Z material.

Since few stars in young clusters have had comprehensive
metallicity analyses, subsequent studies could be
expanded and verified, and including more stars with a larger
range of mass.  Care should be taken
that stars with discrepant metallicities are not thrown out
of the sample and assumed to be 
non-cluster members.  Currently there is no observational evidence 
that metallicities of different mass stars in a given cluster
differ.  Comparison of spectroscopically measured metallicities 
in the Hyades, Pleiades and NGC 2264 between G and F stars find
that the metallicities are the same within 
the observational uncertainties (approximately 0.1 dex)  \citep{king,friel90a}.

Despite the fact that the ISM (along 1kpc line of sight 
absorption lines) appears to have a uniform metallicity
\citep{meyer}, the metallicities of moving groups differ
significantly \citep{friel90a,friel90b},
and the youngest ones seem to have increasingly subsolar
values.  There also are extremely metal rich stars
some of which are not young.
The high metallicity stars are kinematically
related, they tend to be in moving groups or streams \citep{soubiran}.
This suggests that high metallicity stars are formed
in clusters that are uniform in metallicity just like lower metallicity stars.
This may rule out a scenario where even a few percent of stars
have their metallicity increased significantly by the affects
of planetary processes and would instead support a model where
higher metallicity stars are more likely to harbor short
period planets.   However there is good evidence that one
parent star of an extra solar planet has engulfed
planetary material; the detection of Lithium 6 in HD 82943 
\citep{israelian}.  Further study of uniform populations
such as are found in young clusters may be able to 
determine the relative statistical importance of both scenarios.


\acknowledgments

This work could not have been carried out without helpful
discussions with Don Garnett, Caty Pilachowski, Jonathan Lunine,
Ari Laor, Patrick Young,  
Dave Arnett, Adam Burrows, Nadya Gorlova, and Mike Meyer.
I thank Joss de Bruijne for providing me with information on the
high fidelity Hyades sample.
A.~C.~Q.\ gratefully acknowledges hospitality and support 
as a Visitor at the Technion.

\end{document}